\documentclass{ieeeaccess}
\usepackage{cite}
\usepackage{amsmath,amssymb,amsfonts}
\usepackage{newtxtext,newtxmath} 
\usepackage{enumitem}

\AtBeginDocument{%
	\DeclareSymbolFont{letters}{OML}{ntxmi}{m}{it}%
	\SetSymbolFont{letters}{normal}{OML}{ntxmi}{m}{it}%
	\SetSymbolFont{letters}{bold}{OML}{ntxmi}{b}{it}%
	\DeclareSymbolFont{NewLetters}{OML}{ntxmi}{m}{it}%
	\SetSymbolFont{NewLetters}{normal}{OML}{ntxmi}{m}{it}%
	\SetSymbolFont{NewLetters}{bold}{OML}{ntxmi}{b}{it}%
}

\usepackage{algorithmic}
\usepackage{algorithm}
\usepackage{array}
\usepackage{graphicx}
\usepackage{textcomp}
\usepackage{comment}
\usepackage{textcomp}
\usepackage{stfloats}
\usepackage{url}
\usepackage{verbatim}
\usepackage{graphicx}
\usepackage{cite}
\usepackage{color}
\usepackage{soul}

\makeatletter
\AtBeginDocument{\DeclareMathVersion{bold}
	\SetSymbolFont{operators}{bold}{T1}{times}{b}{n}
	\SetSymbolFont{NewLetters}{bold}{T1}{times}{b}{it}
	\SetMathAlphabet{\mathrm}{bold}{T1}{times}{b}{n}
	\SetMathAlphabet{\mathit}{bold}{T1}{times}{b}{it}
	\SetMathAlphabet{\mathbf}{bold}{T1}{times}{b}{n}
	\SetMathAlphabet{\mathtt}{bold}{OT1}{pcr}{b}{n}
	\SetSymbolFont{symbols}{bold}{OMS}{cmsy}{b}{n}
	\renewcommand\boldmath{\@nomath\boldmath\mathversion{bold}}}
\makeatother

\def\BibTeX{{\rm B\kern-.05em{\sc i\kern-.025em b}\kern-.08em
		T\kern-.1667em\lower.7ex\hbox{E}\kern-.125emX}}

\begin{document}
	\history{Received 8 October 2025, accepted 18 October 2025.}
	\doi{10.1109/ACCESS.2025.3624246}
	
	\title{\hl{Automatic Field-of-View Adjustment for \\a View-Expansive Microscope via LSTM-Based Gaze and Pipette Motion Interpretation}}
	\author{\uppercase{Kenta Yokoe}\authorrefmark{1},~\IEEEmembership{Member,~IEEE},
		\uppercase{Takuya Hara}\authorrefmark{2},
		and\\\uppercase{Tadayoshi Aoyama}\authorrefmark{1, 3},~\IEEEmembership{Member,~IEEE}}
	
	\address[1]{Department of Mechanical Systems Engineering, Nagoya University, Nagoya, Aichi, 464-8603, Japan (e-mail: yokoe@nagoya-u.jp, tadayoshi.aoyama@mae.nagoya-u.ac.jp)}
	\address[2]{Department of Micro-Nano Mechanical Science and Engineering, Nagoya University, Nagoya, Aichi, 464-8603, Japan (e-mail: hara@robo.mein.nagoya-u.ac.jp)}
	\address[3]{Center for One Medicine Innovative Translational Research, Gifu University, Yanagido 1-1, Gifu, Gifu 501-1193, Japan}
	\tfootnote{This work was supported by JST CREST, Grant No. JPMJCR20D5, Japan.
		The study protocol was approved by the Ethics Committee of the Graduate School of Engineering, Nagoya University (approval Number: 20-23).
		This is the accepted version of an article published in IEEE Access, vol.~13, pp.~182915--182923, 2025, DOI: 10.1109/ACCESS.2025.3624246. \copyright~2025 The Authors. Open Access under CC BY 4.0 (https://creativecommons.org/licenses/by/4.0/).
	}
	
	\markboth
	{Yokoe \headeretal Automatic Field-of-View Adjustment for \\a View-Expansive Microscope}
	{Yokoe \headeretal Automatic Field-of-View Adjustment for \\a View-Expansive Microscope}
	
	\corresp{Corresponding author: Kenta Yokoe and Tadayoshi Aoyama (e-mail: yokoe@nagoya-u.jp, tadayoshi.aoyama@mae.nagoya-u.ac.jp).}
	
	\sethlcolor{white}
	
	\begin{abstract}
		Intracytoplasmic sperm injection (ICSI) operators frequently adjust the field-of-view (FOV) during procedures, which interrupts workflow and increases procedure time.
		Conventional microscopes require manual objective lens switching and illumination adjustments to achieve different FOV sizes.
		We propose an AI-based automatic FOV adjustment method integrated with a view-expansive microscope.
		This microscope enables the simultaneous acquisition of a large FOV and high-resolution images using a single objective lens through multiview imaging with galvanometer mirrors and high-speed vision, thereby eliminating the need for physical lens exchanges.
		Our method utilizes a long short-term memory (LSTM) model to predict the appropriate FOV size based on real-time analysis of the pipette's position and velocity, combined with the operator's gaze position.
		The AI model is trained using ICSI procedure data from an expert with over five years of micromanipulation experience.
	\hl{Experimental evaluation with novice operators reveals that the proposed automatic FOV adjustment system significantly improves the ICSI procedure speed, reducing the average task completion time from 60.5 to 48.0 s ($p<0.001$). 
		The experiments also demonstrate that this improvement enables novice operators to achieve ICSI working speeds equivalent to those of expert operators.}
	\end{abstract}
	
	
	\maketitle
	
	\section{Introduction}\label{sec:introduction}
	\IEEEPARstart{I}{ntracytoplasmic}
	sperm injection (ICSI) is an assisted reproductive technology (ART) that has experienced significant global growth in recent years \cite{esteves2021Intracytoplasmic}. 
	According to the International Committee for Monitoring Assisted Reproductive Technologies, ICSI currently accounts for approximately 70\% of all ART cycles worldwide \cite{adamson2023ICMART}. 
	This can be attributed to ICSI's capability to address severe male infertility and the improved success rates resulting from the standardization of the procedure \cite{malter2016Micromanipulation,rubino2016ICSI}. 
	Additionally, technical advancements, such as the introduction of laser- and piezo-ICSI, have improved oocyte survival rates and contributed to the global increase in ICSI cases \cite{abdelmassih2002Laserassisted,zander-fox2021PIEZOICSI}.
	
\begin{figure}[!t]
		\centering
		\includegraphics[keepaspectratio=true,width=\linewidth]{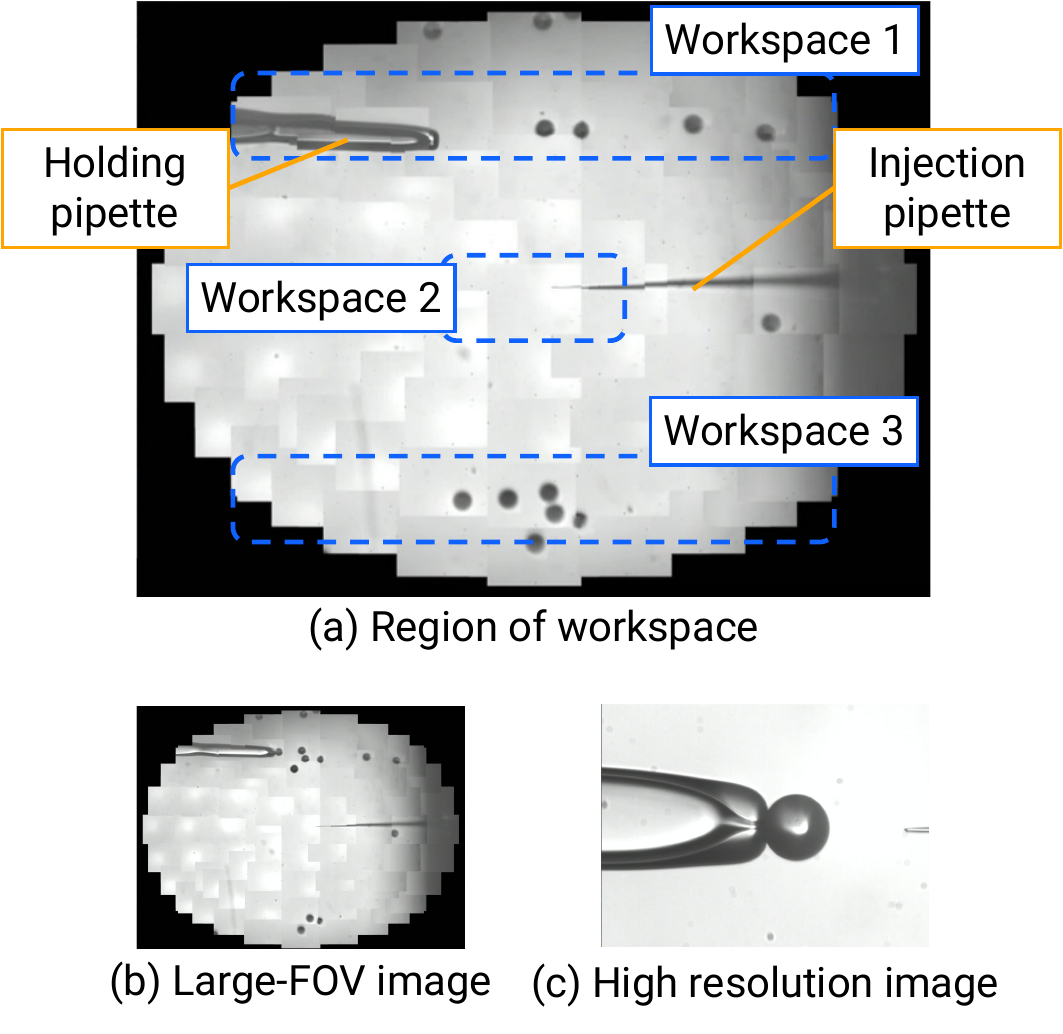}
		\caption{Workspaces of ICSI procedure and image presentation of view-expansive microscope.}
		\label{fig:workspace}
	\end{figure}

	ICSI procedures involve highly complex micromanipulation, requiring precise pipette operations and field-of-view (FOV) adjustments across multiple interfaces.
	As shown in (a) of Fig.~\ref{fig:workspace}, ICSI is conducted in three distinct workspaces: 
	\begin{description}[style=multiline, leftmargin=2.25cm]
		\item[Workspace 1:] Area where oocytes are placed before injection
		\item[Workspace 2:] Injection area
		\item[Workspace 3:] Area where oocytes are released after injection
	\end{description}
	The ICSI procedure consists of three main processes performed repeatedly: (1) Moving the oocyte from Workspace 1 to Workspace 2 using a holding pipette;
	(2) injecting sperm into the target oocyte with an injection pipette in Workspace 2; and 
	(3) transferring the treated oocyte from Workspace 2 to Workspace 3. Operators must adjust the FOV and brightness to perform these tasks smoothly.
	For example, task (1) is performed at a magnification level that allows the target oocyte to be held with a holding pipette (typically using a medium-magnification lens). Task (2) requires high-resolution observation for precise manipulation with the injection pipette (typically using a high-magnification lens), and task (3) involves releasing the oocyte and returning to Workspace 1 under a large FOV (typically using a low-magnification lens). 
	As changing the objective lens magnification affects brightness, operators must adjust the brightness accordingly.
	Switching the objective lens using a revolver can also cause operators to lose sight of the oocytes.
	These frequent manual adjustments to objective lenses and brightness settings disrupt workflow, increase procedure time, and impact the success rate and reproducibility of ICSI.
	Consequently, the complexity of ICSI procedures necessitates extensive training to master it \cite{costa-borges2023First,durban2016Area,lu2011Robotic}.
	
	Our research group has previously developed a view-expansive microscope that enables the simultaneous presentation of large-FOV images ((b) in Fig.~\ref{fig:workspace}) and high-resolution images ((c) in Fig.~\ref{fig:workspace}) using galvanometer mirrors and high-speed vision \cite{aoyama2023Micromanipulation}. 
	This microscope eliminates the need for objective lens changes and brightness adjustments during ICSI procedures, allowing operators to achieve faster micromanipulation than with conventional microscopes without losing sight of the oocyte.
	However, this microscope presents operators with vast amounts of visual information, potentially exceeding human data-processing capabilities. 
	Additionally, operators must manually determine the FOV size of the displayed image, which diverts operators' attention from the primary manipulation task.
	
\begin{figure}[!t]
		\centering
		\includegraphics[keepaspectratio=true,width=.85\linewidth]{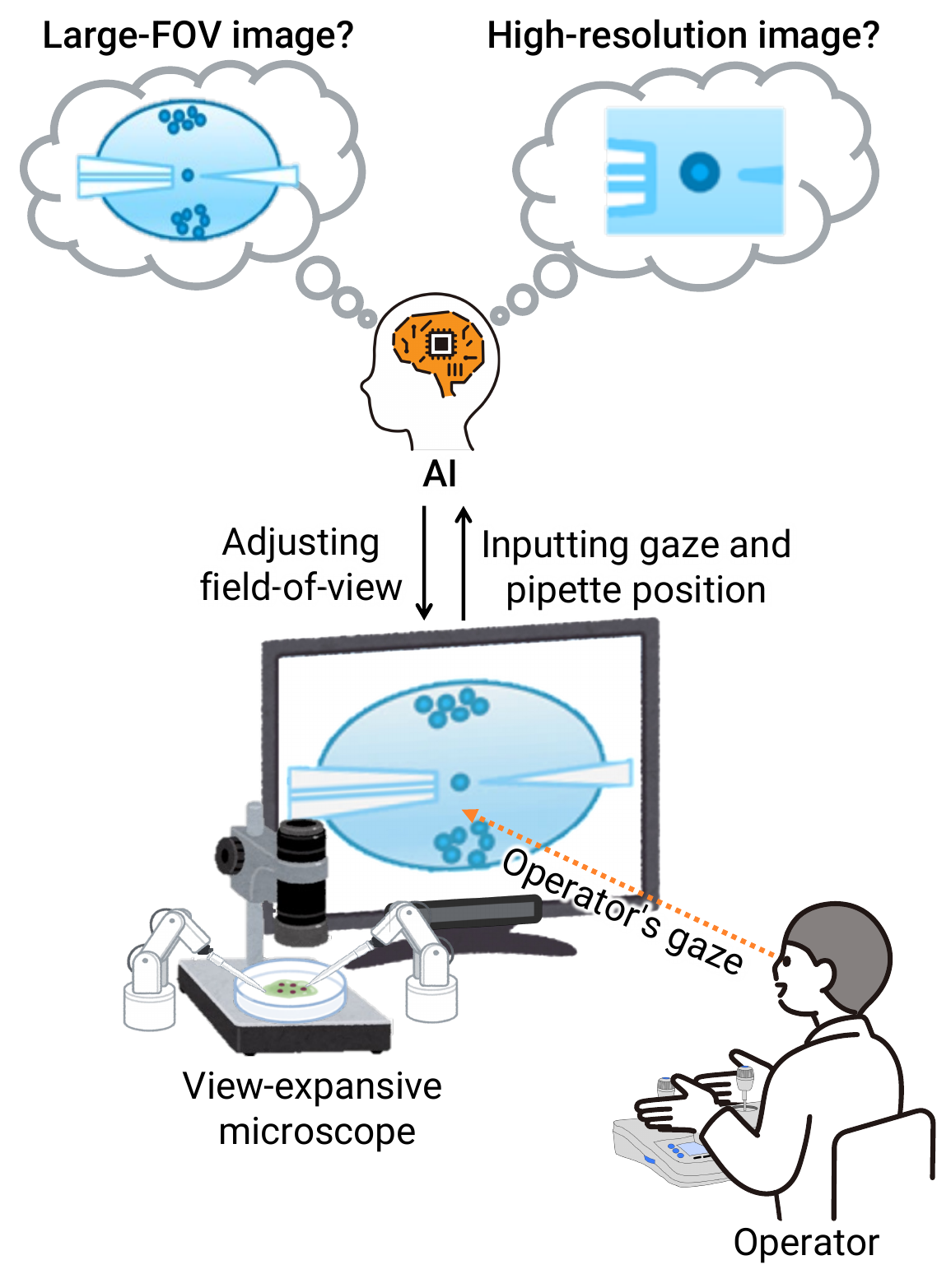}
		\caption{Schematic of proposed system. The AI model automatically adjusts FOV of displayed image based on operator's gaze and pipette's position.}
		\label{fig:concept}
	\end{figure}

	To address these limitations, we proposed an AI-based automatic FOV adjustment method integrated with a view-expansive microscope.
	Figure \ref{fig:concept} presents a schematic of the proposed system.
	Our proposed system employs an LSTM model that predicts the appropriate FOV size based on real-time analysis of the pipette's position and velocity, combined with the operator's gaze position. 
	The operator's intention can be estimated by incorporating gaze movements as input to the AI.
	The AI model was trained using ICSI procedure data from an expert with over five years of micromanipulation experience. The proposed method eliminates the need for operators to manually change the FOV size, allowing them to maintain a continuous focus on primary manipulation tasks without interruption.
	To evaluate the proposed system, we conducted an experiment with novice operators who performed ICSI-mimicking tasks using microbeads. 
	The experimental results demonstrated significant improvements in ICSI speed, enabling novices to achieve ICSI performance at the same speed as experts.
	The remainder of this  paper is organized as follows:
	Section \ref{sec:introduction} provides the introduction and motivation for this study.
	Section \ref{sec:related_work} reviews related work on automation and support systems for ICSI, as well as gaze-based intention recognition for human-robot interactions.
	Section \ref{sec:view-expansive_microscope_system_with_gaze_tracking_functionality} describes the view-expansive microscope system with gaze-tracking functionality. 
	Section \ref{sec:fov_adjustment_algorithm_using_ai} presents an AI-based FOV adjustment algorithm that utilizes LSTM. 
	Section \ref{sec:speed_evaluation_of_icsi_procedure} evaluates the system's performance through speed evaluation experiments on ICSI procedures with novice and expert operators. 
	Section \ref{sec:discussion} discusses the implications of our findings and the limitations of the proposed approach. 
	Finally, Section \ref{sec:conclusion} concludes the paper and outlines future research directions.
	
	\section{Related Work}\label{sec:related_work}
	Numerous studies have been conducted on the automation and simplification of micromanipulation procedures, including ICSI \cite{malter2016Micromanipulation}.
	Among these, laser-ICSI and piezo-ICSI are representative technologies that have been implemented at the clinical level and are widely used \cite{choi2011Efficiency,caddy2023PIEZOICSI}.
	Additionally, research has demonstrated the effectiveness of using microfluidic technology for sperm and oocyte selection \cite{aderaldo2023Does,olatunji2022Review,weng2018Onchip,fang2023Microfluidic}. 
	Most of these technologies aim to improve embryo success rates and the speed of ICSI by providing supplementary assistance to existing procedures rather than modifying the established ICSI workflow.
	However, several studies have focused on automating the ICSI procedure \cite{lu2011Robotic}.
	These studies aimed to replace the complex ICSI procedures performed by humans with robotic systems, reporting automation of tasks, including sperm immobilization \cite{zhang2019Robotica}.
	Although these studies have automated some tasks within the ICSI procedure, they have not achieved complete automation, still necessitating human operators.
	
	Therefore, instead of pursuing full automation, research on remote ICSI has been conducted to maximize the utilization of existing ICSI experts \cite{mendizabal-ruiz2025digitally,mendizabal2024O185}.
	While these systems require on-site operators, the studies enable ICSI experts to perform challenging tasks remotely, allowing for high-quality ICSI even in regions with few ICSI experts.
	\hl{These studies utilized conventional microscopes and ICSI interfaces, resulting in no reduction in the burden on experts.}
	Additionally, efforts have been made to reduce the complexity of ICSI by enhancing microscope systems and interfaces, thereby increasing the number of experts \cite{fujishiro2021Microinjectiona,yokoe2022immersive,mori2024RealTime,costa-borges2023First}.
	These studies have simplified several skilled techniques required for ICSI by assisting operators with machines and AI, thus improving the speed and accuracy of the procedures.
	\hl{However, these systems provide reactive assistance based on current operator actions, rather than proactive support that anticipates operator intentions.}
	
	In collaborative systems involving humans, machines, and AI, the critical factor is delivering support that aligns with human operations \cite{trombetta2021Variable}.
	To estimate human intention, many studies have utilized gaze movements in human-robot interactions \cite{pan2024Gazebased,belardinelli2024GazeBased}.
	These studies have achieved smooth human-robot interaction systems by estimating human intentions through gaze information.
\hl{Additionally, gaze-based interface control has been explored across multiple domains including user interfaces} \cite{stellmach2012Designing}\hl{, surgical robotics} \cite{zhang2025GazeScope,fujii2018Gaze}\hl{, automotive applications} \cite{prabhakar2020Interactive,lee2020Automotive}\hl{, and virtual reality (VR) and augmented reality (AR) systems} \cite{konrad2020GazeContingent}\hl{. 
	These studies have achieved smooth human-robot interaction, such as FOV adjustment, by estimating human intentions through gaze information. However, these approaches employ the gaze as an explicit input modality, requiring operators to consciously perform specific gaze patterns or gestures to trigger system responses. Furthermore, these approaches have been applied primarily to macro-scale robot interactions and have not been adapted to micro-scale environments. Although gaze-based intention recognition has shown significant effectiveness in human-robot interaction systems, applications in ICSI support systems are limited. Therefore, this study addresses this application gap by introducing gaze-based intention recognition into ICSI support systems. In contrast to prior explicit gaze-based control systems, the proposed approach employs AI to interpret natural task-related gaze behavior implicitly, eliminating the need for conscious gaze control and enabling truly automatic FOV adjustment that anticipates operator intentions.}

	Specifically, we developed an LSTM-based model that integrates real-time gaze tracking with pipette position to predict the appropriate FOV size during ICSI procedures. 
	By combining gaze-based intention recognition, our approach enables human-centered ICSI assistance that anticipates human needs rather than merely responding to explicit commands.
	
	\section{View-Expansive Microscope System with \hl{AI-Based FOV Adjustment}}
	\subsection{\hl{Hardware Configuration}}\label{sec:view-expansive_microscope_system_with_gaze_tracking_functionality}
	\begin{figure}[!t]
		\centering
		\includegraphics[keepaspectratio=true,width=\linewidth]{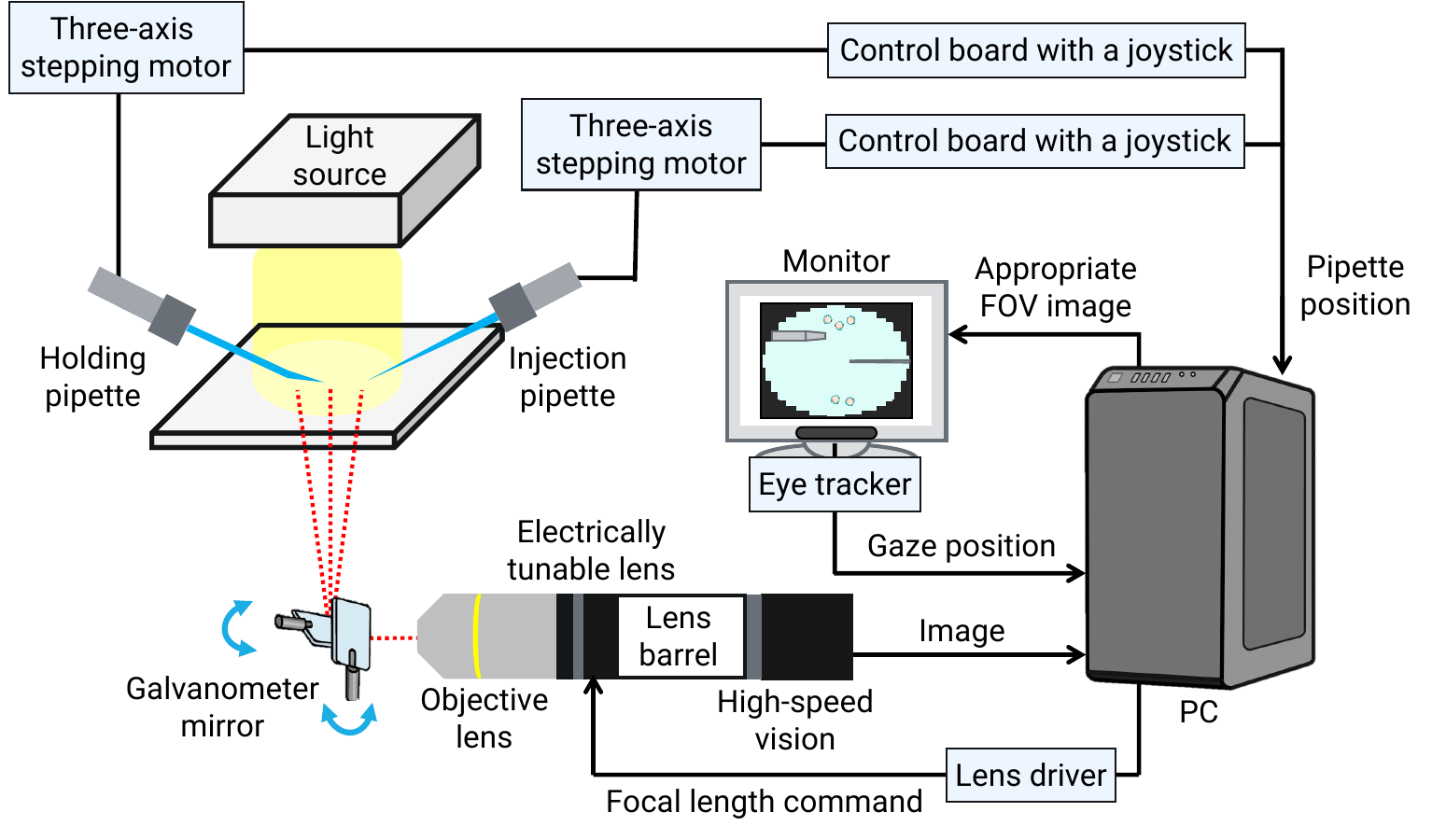}
		\caption{Configuration of proposed system. }
		\label{fig:system}
	\end{figure}
	\begin{figure}[!t]
		\centering
		\includegraphics[keepaspectratio=true,width=\linewidth]{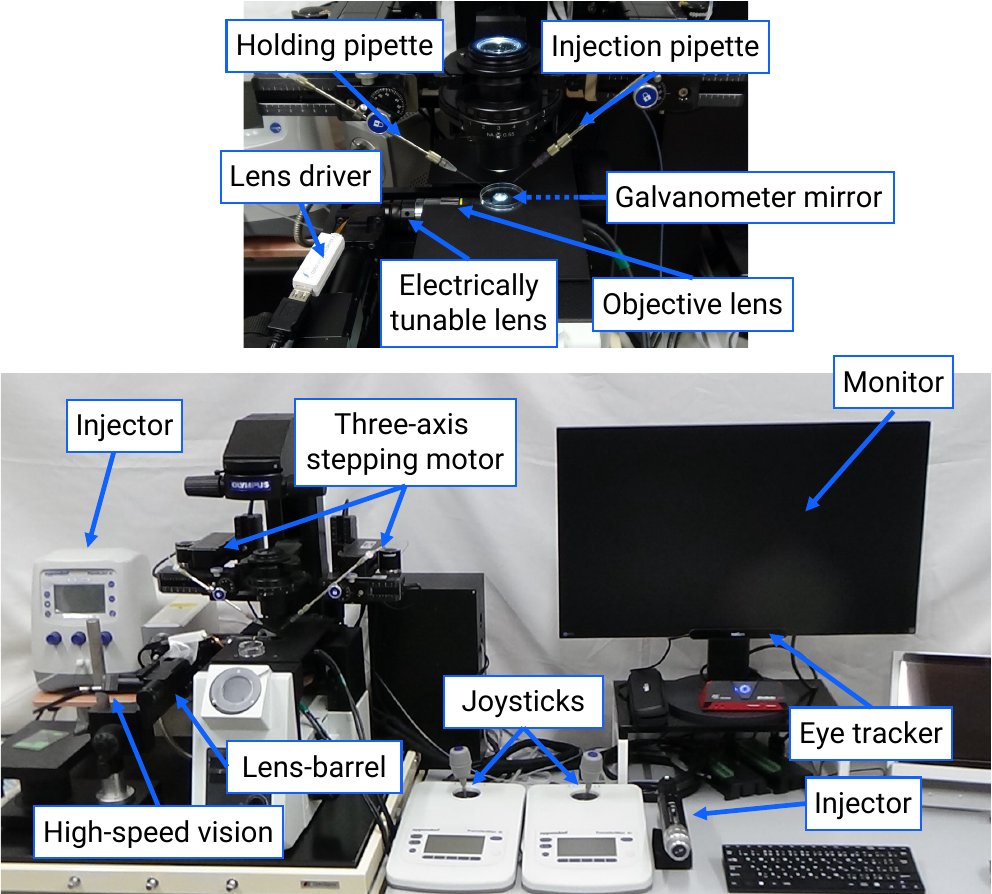}
		\caption{Overview of proposed system.}
		\label{fig:overview}
	\end{figure}
	Figures \ref{fig:system} and \ref{fig:overview} present the configuration and overview of the proposed system, respectively.  
	The proposed system is based on the view-expansive microscope developed in a previous study \cite{aoyama2023Micromanipulation}. 
	\hl{In this study, gaze tracking functionality was added to the view-expansive microscope to understand the intention of the operator from gaze movements. The view-expansive microscope achieves high-resolution imaging through optical means (galvanometer mirrors and an electrically tunable lens) rather than computational super-resolution, ensuring real-time performance that is essential for ICSI procedures.}
	The proposed system comprises an inverted microscope (IX73, OLYMPUS), a high-speed vision (MQ003MG-CM, Ximea), an electrically tunable lens (EL-10-30-C-VIS-LD-MV, Optotune), a lens driver (Lens Driver 4, Optotune), a two-axis galvanometer mirror (6210HSM 6 mm 532 nm, Cambridge Technology), a light source unit (LA-HDF158AS, Hayashi Repic Co., Ltd.), a control computer (OS Windows 10 Pro 64-bit, CPU Intel (R) Core (TM) i9-10859K 3.60 GHz, memory 32 GB RAM, GPU NVIDIA GeForce RTX), a monitor (FlexScan EV2781, EIZO), a D/A board (PEX-340416, interface), an A/D board (PEX-321216, interface), two microinjectors (FemtoJet 4i and CellTram 4r Air, Eppendorf), two micromanipulators (TransferMan4r, Eppendorf), a screen-based eye tracker (Tobii Pro Nano, Tobii), and an objective lens (LWD 95 mm, Mitsutoyo). 
	The magnification of the objective lens was 10$\times$, with a working distance of 33 mm. The pixel pitch of the high-speed vision was 7.4 $\mathrm{\mu}$m, enabling image capture at a resolution of 640$\times$480 pixels and a maximum frame rate of 500 fps.
	This system configuration facilitates the simultaneous acquisition of high-resolution, large-FOV images with automatic FOV adjustment based on gaze tracking.
	
	\subsection{\hl{Algorithm Design}}\label{sec:fov_adjustment_algorithm_using_ai}
	The proposed system employs an AI model to predict the appropriate FOV size of the view-expansive microscope based on gaze and pipette movements.
	\begin{table}[!t]
		\caption{\hl{AI model structure. }}
		\label{table:model_config}
		\centering
		\begin{tabular}{c|c}
			\textbf{Parameter} & \textbf{Value} \\
			\hline
			Input & 8 nodes \\
			LSTM & 8 nodes (1 layer) \\
			Output & 1 node \\
			Optimizer & RMSprop \\
			Loss Function & Mean Squared Error (MSE) \\
			Batch Size & 8 \\
			Epochs&85\\
		\end{tabular}
	\end{table}
	\hl{Table} \ref{table:model_config} \hl{shows the structure of the AI model used in the proposed method.}
\hl{This study applies the LSTM architecture, considering system efficiency and real-time performance. ICSI involves handling irreplaceable biological material where procedural failures cannot be reversed, requiring extremely high system reliability and local operation without internet connectivity for patient privacy protection and connection stability. Transformer-based architectures were excluded due to their computational requirements and infrastructure dependencies. LSTM and gated recurrent units (GRUs)  offer comparable real-time performance that is suitable for clinical requirements. Given this sufficient inference speed, LSTM was selected for its superior accuracy in capturing complex temporal dependencies} \cite{shiri2024Comprehensive}\cite{shi2025comparative}\hl{, prioritizing modeling capacity over marginal speed improvements. The separate forget and input gates of LSTM enable the model to maintain long-term context during three-workspace transitions while remaining responsive to immediate dynamics. Therefore, the proposed AI model operates locally without internet connectivity, eliminating communication-related latency and ensuring patient privacy.}
	The AI input consists of eight data points updated at a frequency of 20 Hz: 
	\begin{enumerate}
		\setlength{\parskip}{0cm} 
		\setlength{\itemsep}{0cm} 
		\item[1)] Holding pipette position on the monitor ($x_{\mathrm{h}}, y_{\mathrm{h}}$) 
		\item[2)] Holding pipette velocity on the monitor ($\dot{x}_{\mathrm{h}}, \dot{y}_{\mathrm{h}}$)
		\item[3)] Operator's gaze position on the monitor ($x_{\mathrm{g}}, y_{\mathrm{g}}$)
		\item[4)] Deviation of gaze position from holding pipette position ($x_{\mathrm{g}} - x_{\mathrm{h}}, y_{\mathrm{g}} - y_{\mathrm{h}}$)
	\end{enumerate}
\hl{While vision-based AI systems using deep learning have demonstrated success in human-machine interaction applications } \cite{xing2025Visual,xing2025Intelligent}
\hl{, these methods often lack real-time performance due to high computational costs. To achieve the real-time responsiveness required for ICSI, the proposed system was designed using pipette position and velocity data rather than image-based features. 
	The holding pipette position/velocity and gaze position were selected as input features because the holding pipette is actively manipulated throughout all ICSI phases and gaze information effectively captures operator attention. The injection pipette position/velocity was excluded because it exhibits minimal movement compared to the holding pipette, and the oocyte position/velocity was excluded because oocytes are manipulated by the holding pipette, making this information redundant. Therefore, the selected eight-dimensional features are sufficient for FOV prediction.} 
	These inputs are normalized to a range of 0.0 to 1.0 before being fed into the AI model.
	The output of the AI is the appropriate FOV of the image.
	The appropriate FOV parameter, $F_{\mathrm{pred}}$, is an output value between 0.0 and 1.0.
	For the training data, we utilized data from 14 trials (11,520 steps) of ICSI procedures performed by an operator familiar with ICSI procedures. 
	The validation data were derived from three trials (2,768 steps) of the ICSI procedure conducted by the same operator.
	\hl{The AI model was trained using data from a single expert to avoid the multimodality problem inherent in ICSI procedures. Different experts and clinics employ varying operational techniques, pipette manipulation styles, and FOV preferences. Combining such diverse training data could result in conflicting learning targets that degrade model performance. Single-expert training was strategically selected to ensure consistent learning objectives for the LSTM model.
	Each ICSI procedure in the training and validation datasets represents a complete procedure performed on a single oocyte, encompassing the entire three-workspace workflow described in Section} \ref{sec:introduction}\hl{: oocyte holding and positioning in Workspace 1, sperm injection in Workspace 2, and treated oocyte release in Workspace 3. 
	The actual expert operational data cannot be shared publicly due to proprietary concerns regarding individual identification and unauthorized replication of specialized micromanipulation techniques.}
	We employed PyTorch for the AI model.
	The coefficient of determination was 0.91.
	\hl{The AI inference time is 3.9-6.0 ms, which is sufficient to avoid affecting the response time of the view-expansive microscope.}
	
	Based on the appropriate FOV parameter $F_{\mathrm{pred}} \in [0, 1]$ output by the proposed AI model, the view-expansive microscope adjusts the FOV of the microscopic image.
	The width of the image displayed to the operator, $L_{\mathrm{d}}$, is determined by the actual FOV parameter $F$ as follows:
	\begin{align}\label{eq:image_width}
		L_{\mathrm{d}} = F L_{\mathrm{s}} + (1 - F) L_{\mathrm{e}}, 
	\end{align}
	where $L_{\mathrm{s}}$ represents the width of the highest-resolution image captured by our view-expansive microscope, and $L_{\mathrm{e}}$ represents the maximum expanded FOV. 
	When $F = 1$, the system displays the highest-resolution image ($L_{\mathrm{d}} = L_{\mathrm{s}}$). 
	When $F = 0$, the system displays the maximum FOV image ($L_{\mathrm{d}} = L_{\mathrm{e}}$). 
	In our implementation, $L_{\mathrm{s}}$ corresponded to 500 $\mathrm{\mu}$m, while $L_{\mathrm{e}}$ could reach up to 1.5 mm.
	The FOV parameter $F$ was controlled to track the AI prediction $F_{\mathrm{pred}}$ as follows:
	\begin{align}\label{eq:rate_limit}
		\dot{F} = \text{sgn}(F_{\mathrm{pred}} - F) \cdot \min(|F_{\mathrm{pred}} - F|, \dot{F}_{\mathrm{max}}), 
	\end{align}
	where $\dot{F}_{\mathrm{max}}$ represents the maximum speed of FOV change required to prevent abrupt visual transitions, which can cause operator disorientation. 
	Specifically, rapid FOV changes interfere with the operator's observation, leading to a loss of sight of the pipettes and oocytes. 
	Therefore, operators must frequently re-acquire pipettes and oocytes, which degrades manipulation precision and extends the duration of the ICSI procedure.
	Such issues are commonly observed in traditional microscopes when operators manually change the objective lenses during delicate micromanipulation procedures.
	To adjust the FOV size using Equations \eqref{eq:image_width} and \eqref{eq:rate_limit}, the center of the FOV ($x_{\mathrm{c}}, y_{\mathrm{c}}$) is determined as follows:
	\begin{align}\label{eq:center_of_view}
		\begin{cases}
			x_{\mathrm{c}} = x_{\mathrm{h}}+r_{\mathrm{o}} \\
			y_{\mathrm{c}} = y_{\mathrm{h}}
		\end{cases},
	\end{align}
	where $x_{\mathrm{h}}$ and $y_{\mathrm{h}}$ represent the position of the holding pipette, and $r_{\mathrm{o}}$ represents the radius of the oocyte.
	By adjusting the FOV centered at the position defined by Equation \eqref{eq:center_of_view}, operators can perform smooth ICSI without losing sight of the holding pipette or oocyte.
	
	\section{Speed Evaluation of ICSI Procedure}\label{sec:speed_evaluation_of_icsi_procedure}
	\subsection{Experimental Setup}\label{subsec:experimental_setup}
	\begin{figure}[!t]
		\centering
		\includegraphics[keepaspectratio=true,width=.8\linewidth]{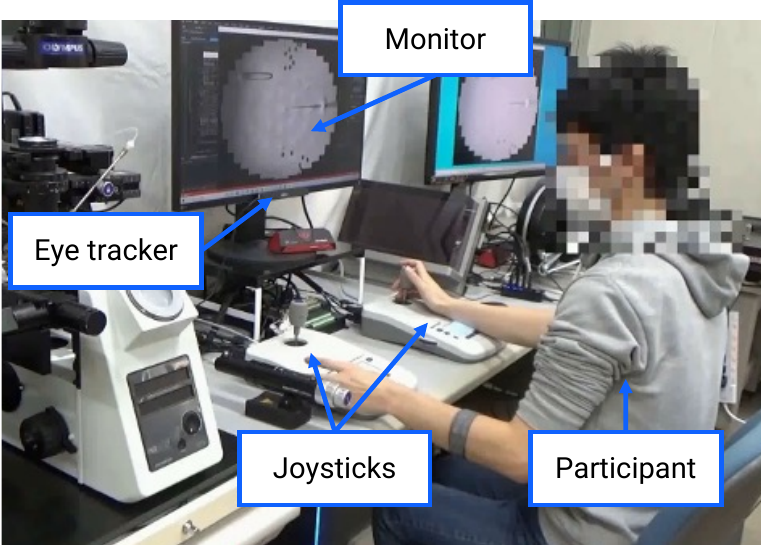}
		\caption{Overview of a participant during experiment.}
		\label{fig:participant}
	\end{figure}
	To evaluate the effectiveness of the proposed system, we conducted an experiment using 100 $\mathrm{\mu}$m microbeads, which are similar in size to human oocytes. 
	The experiment was designed to simulate the ICSI procedure described in Section \ref{sec:introduction}. 
	Participants were required to perform the following ICSI procedure: pick up a microbead from Workspace 1, move it to Workspace 2, touch an injection pipette tip to the microbead, and then move the microbead to Workspace 3. 
	
	Participants performed this procedure using six microbeads.
	The experiment was conducted under two conditions: with a view-expansive microscope featuring automatic FOV adjustment and with a view-expansive microscope without automatic FOV adjustment. 
	In the absence of automatic FOV adjustment, participants adjusted their FOV using a mouse wheel.
	\hl{This experimental design isolates the effect of automatic FOV adjustment by comparison with manual adjustment within the same microscope system. Comparisons with conventional microscopes or other gaze-based FOV systems were not included because the advantages of a view-expansive microscope over a conventional microscope have already been demonstrated} \cite{aoyama2023Micromanipulation}\hl{, and existing gaze-based systems employ fundamentally different operational paradigms requiring explicit gaze control. Instead, expert performance was included as a reference benchmark.}
	Figure \ref{fig:participant} presents an overview of the participants during the experiment.
	Six individuals with no prior experience in micromanipulation participated
	in the study.
	Before the experiment, participants received an explanation of the experimental procedure and view-expansive microscope system. 
	Informed consent was obtained from all participants before the experiment commenced.
	Participants then completed a ten-minute practice session to familiarize themselves with the task and the view-expansive microscope.
	After the practice session, participants performed the experimental task using six microbeads. 
	The order of experimental conditions was counterbalanced for each participant.
	Additionally, an expert with over five years of micromanipulation experience performed the same experimental task using a conventional microscope system that they were familiar with.
	We evaluated the proposed system by measuring the time participants took to complete the experimental task.
	
	\subsection{Experimental Results}\label{subsec:experimental_results}
	\begin{figure*}[!t]
		\centering
		\includegraphics[width=.95\linewidth]{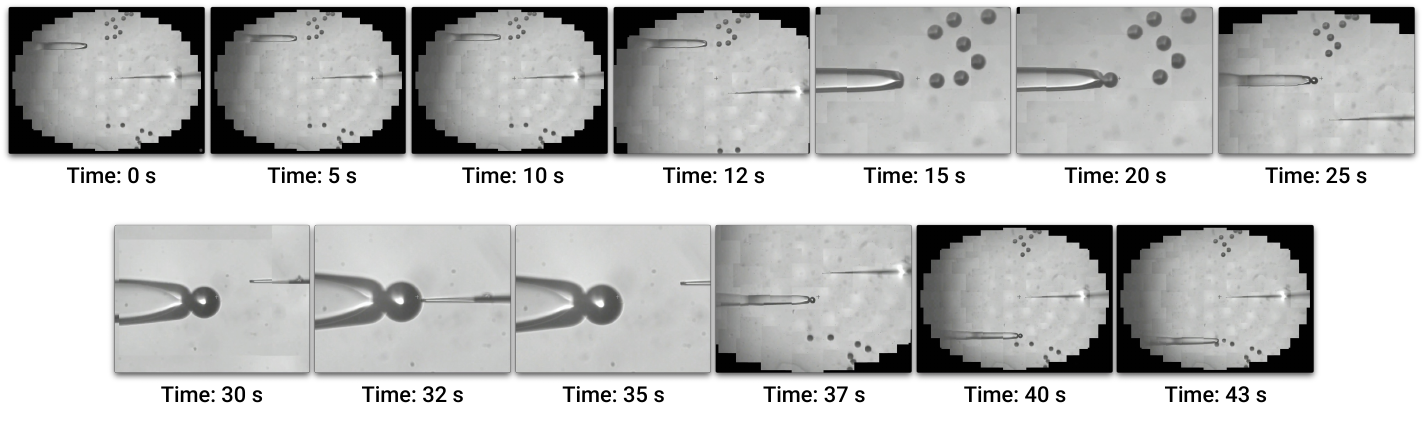}
		\caption{Time series of images presented to operator during use of proposed system.}
		\label{fig:time_series}
	\end{figure*}
	Figure \ref{fig:time_series} shows the time series of the images presented to the operator during the experiment with the proposed system.
	At the beginning of the experimental task, participants moved the holding pipette to Workspace 1 in Fig. \ref{fig:workspace} to hold a microbead (0--10 s in Fig. \ref{fig:time_series}). 
	Upon the arrival of the holding pipette in Workspace 1, the proposed system adjusted the FOV (12–15 s in Fig. \ref{fig:time_series}). 
	Each participant held one microbead (15--20 s in Fig. \ref{fig:time_series}). 
	Subsequently, while transporting the microbead to Workspace 2, the FOV was expanded using the proposed system (20--25 s in Fig. \ref{fig:time_series}). 
	During the injection in Workspace 2, the proposed system presented high-resolution images (25--35 s in Fig. \ref{fig:time_series}). 
	After completing the injection, the proposed system presented a large FOV image to the participant, who then moved the microbead to Workspace 3 (37--43 s in Fig. \ref{fig:time_series}). 
	The participants ejected the microbead (43--48 s in Fig. \ref{fig:time_series}).
	This confirms that the proposed system can adjust the FOV according to gaze and pipette movements.
	
	\begin{figure}[!t]
		\centering
		\includegraphics[keepaspectratio=true,width=\linewidth]{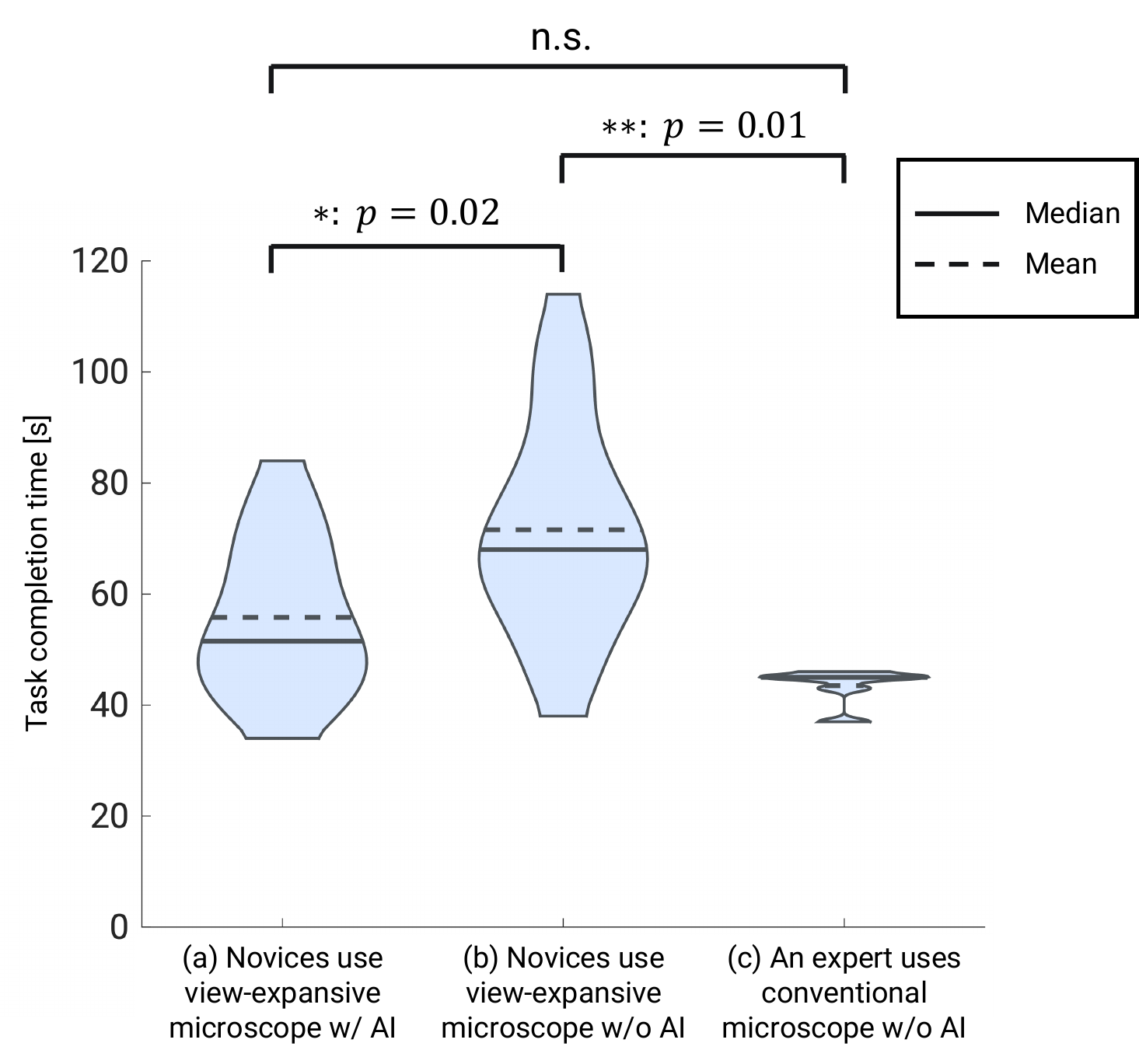}
		\caption{Violin plot of task completion time for each condition. Solid line represents median value, while dashed line represents mean value.}
		\label{fig:result}
	\end{figure}
	Figure \ref{fig:result} displays a violin plot of the task completion time for each condition.
	The solid line indicates the median values, and the dashed line represents the mean values.
	The median time taken by novices using the view-expansive microscope with and without the proposed automatic FOV adjustment was 51.5 s and 68 s, respectively.
	Meanwhile, the median time taken by experts using the conventional microscope system was 45 s. 
	Results from the Kruskal–Wallis test with Bonferroni correction indicated that novices using the proposed system (i.e., the view-expansive microscope with automatic FOV adjustment) showed a significant difference compared to those using the view-expansive microscope without automatic FOV adjustment ($p = 0.02$). There was no significant difference, however, when compared to the conventional microscope. 
	These results suggest that the proposed automatic FOV adjustment can enhance the speed of micromanipulation with a view-expansive microscope, enabling novices to perform micromanipulation at a speed comparable to that of experts using a conventional microscope.
	
	\begin{table}[!t]
		\centering
		\caption{Median time of experimental task for each participant under each condition.}
		\label{tab:participant}
		\begin{tabular}{c|cc}
			Participant & \begin{tabular}[c]{@{}c@{}}(a) Automatic FOV\\ adjustment [s]\end{tabular} & \begin{tabular}[c]{@{}c@{}}(b) Manual FOV\\ adjustment [s]\end{tabular} \\ \hline
			A & 73.0 & 101.0 \\
			B & 44.0 & 64.5  \\
			C & 56.0 & 81.0  \\
			D & 58.5 & 63.0  \\
			E & 42.5 & 47.0  \\
			F & 65.0 & 69.5 \\ \hline
			All & 51.5 & 68.0 \\
		\end{tabular}
	\end{table}
	Table \ref{tab:participant} lists the median time of the experimental task for each participant under each 
	condition (A--F represent the participants).
	This result demonstrates that the proposed method reduced the median task completion time for all participants.
	Furthermore, the median values for participants B and E using the proposed system were lower than those of the experts. 
	This indicates that certain novices performed ICSI procedures more rapidly than experts using a conventional microscope.
	
	\section{Discussion}\label{sec:discussion}
	The experimental results demonstrated that integrating the proposed AI-based FOV adjustment into a view-expansive microscope improved the speed of ICSI procedures for novice operators. 
	This speed is comparable to that of an expert performing ICSI with a conventional microscope.
	This finding is attributed to the automatic FOV adjustment, which eliminates the need for operators to manually adjust the magnification settings.
	When using a conventional microscope, the operator must rotate the revolver to switch the magnification of the objective lens. 
	At this point, it is necessary to adjust the brightness of the microscopic image in accordance with the change in objective lens magnification.
	In contrast, a view-expansive microscope can present images ranging from a large FOV to high resolution using a single objective lens. 
	Therefore, there is no need to rotate the revolver for FOV changes, and brightness adjustment is not required.
	However, FOV adjustments must be made using an interface other than a revolver (such as a mouse wheel in the experiments).
	The proposed automatic FOV adjustment removes the need for an interface for FOV 
	adjustments, allowing the operator to perform experimental tasks without removing their hands from the joystick used for pipette manipulation.
	This elimination of interface-switching is believed to enhance the speed of the ICSI procedure.
	To further enhance operational speed in human-machine systems, it may be necessary to focus not only on simplifying interfaces but also on reducing the number of interfaces that operators must manage. 
	
	The experimental results showed individual differences among the subjects. While participants B and E achieved ICSI procedure speeds exceeding those of the expert, participants D and F showed only a 4.5 s time reduction compared to manual FOV adjustment and remained slower than the expert.
\hl{Nevertheless, all participants showed performance improvements, and the median task completion time for novice operators using the proposed system (51.5 s) was statistically equivalent to expert performance using conventional microscopes (45 s, p > 0.05), successfully achieving the primary research objective of enabling novice operators to perform ICSI procedures at expert-level speeds.}
	The proposed AI model was trained by a single expert, and as a result, it presents the operator with the expert's preferred FOV size during ICSI procedures. 
	Consequently, while the FOV size may be suitable for some operators, it may be too large or too small for others.
	Furthermore, there may be individual differences in gaze movement patterns during ICSI, and the proposed AI is likely to achieve smooth FOV changes for novices whose gaze movement patterns resemble those of experts.
\hl{Previous studies have demonstrated such individual differences in human gaze patterns }\cite{packard2023Stable,broda2024Individual}. 
\hl{The performance variations among participants may reflect these individual gaze pattern differences. Using a single-expert training dataset limits the ability of the system to accommodate diverse individual gaze patterns, indicating the need for future AI models that can compensate for individual differences in FOV preferences and gaze movement patterns.}
	These individual differences in FOV preferences and gaze movement patterns account for the variance in task completion times observed across participants.
	Such individual differences in the effectiveness of the proposed system are anticipated to diminish by increasing the number of experts used for training.
	However, experts may also differ in their preferred FOV sizes and gaze movement patterns---referred to as multimodality---indicating that alternative methods capable of handling multimodality may be necessary when training with data from multiple experts.
	
	\hl{The proposed system employs a static pre-trained LSTM model to maintain constant performance once deployed. This approach prioritizes stability and reliability, which are paramount for medical applications. However, this model cannot be directly applied to different microscope environments or tasks and would require retraining for new operational conditions. Furthermore, developing systems capable of handling multiple tasks would necessitate architectural modifications to accommodate diverse datasets and varied operational tasks.}
	
\hl{The proposed system employs the Tobii Pro Nano eye tracker for gaze acquisition. Like all screen-based eye trackers, this device requires per-operator calibration, which is time-consuming and does not always succeed on the first attempt. Moreover, calibration accuracy varies between sessions, introducing variability in gaze measurement precision. These calibration requirements represent a significant practical burden that must be addressed to improve system usability for clinical settings. Additionally, the gaze acquisition accuracy and drift characteristics of the eye tracker fundamentally limit the overall system performance. The eye tracker can experience drift during extended use due to head movements or lighting changes, potentially requiring recalibration between procedures in extended clinical sessions. These hardware-dependent limitations constrain system performance regardless of AI model improvements. Future enhancements require not only AI algorithm refinement but also advances in eye tracking hardware, such as improved tracking robustness, calibration-free methods, or more reliable gaze acquisition technologies.}
	
	\section{Conclusion}\label{sec:conclusion}
	In this study, we developed an AI-based automatic FOV adjustment system for view-expansive microscope to enhance the speed of ICSI procedures. 
	The proposed system utilizes an AI model that predicts the appropriate FOV size based on the position and velocity of the holding pipette, as well as the operator's gaze position on the monitor. 
	The AI model was trained using data from an operator familiar with ICSI procedures.
	Experimental results demonstrated that novice operators using the proposed automatic FOV adjustment system achieved significantly shorter task completion times than those using manual FOV adjustment.
	Moreover, this performance was statistically equivalent to that of an expert operator using a conventional microscope, indicating that the proposed system enables novices to perform micromanipulation at expert-level speeds.
	The key contribution of this study lies in eliminating the need for manual FOV adjustments during ICSI procedures. 
	By automatically adjusting the FOV according to the ICSI procedure, operators can maintain continuous operation of the pipettes without interruption, thereby enhancing operational speed. 
	This finding suggests that reducing the number of interfaces requiring human operation is crucial for improving the performance of human-machine systems.
	
\hl{Future studies should develop algorithms capable of handling multimodality across different operational styles of experts while incorporating adaptive fine-tuning mechanisms for individual operators. Additionally, addressing the inherent procedural variations across different clinics and embryologists will enhance the adaptability and personalization capabilities of the system across a broader range of operators. Moreover, advances in eye tracking hardware, such as calibration-free methods or improved tracking robustness, are essential for reducing the practical burden of system setup and maintenance in clinical environments. These hardware improvements, combined with algorithmic advances, will facilitate broader clinical adoption of automatic FOV adjustment systems.}
	\bibliography{AutoZooming1.bib}
	\bibliographystyle{IEEEtran}
	
	\begin{IEEEbiography}[{\includegraphics[width=1in,height=1.25in,clip,keepaspectratio]{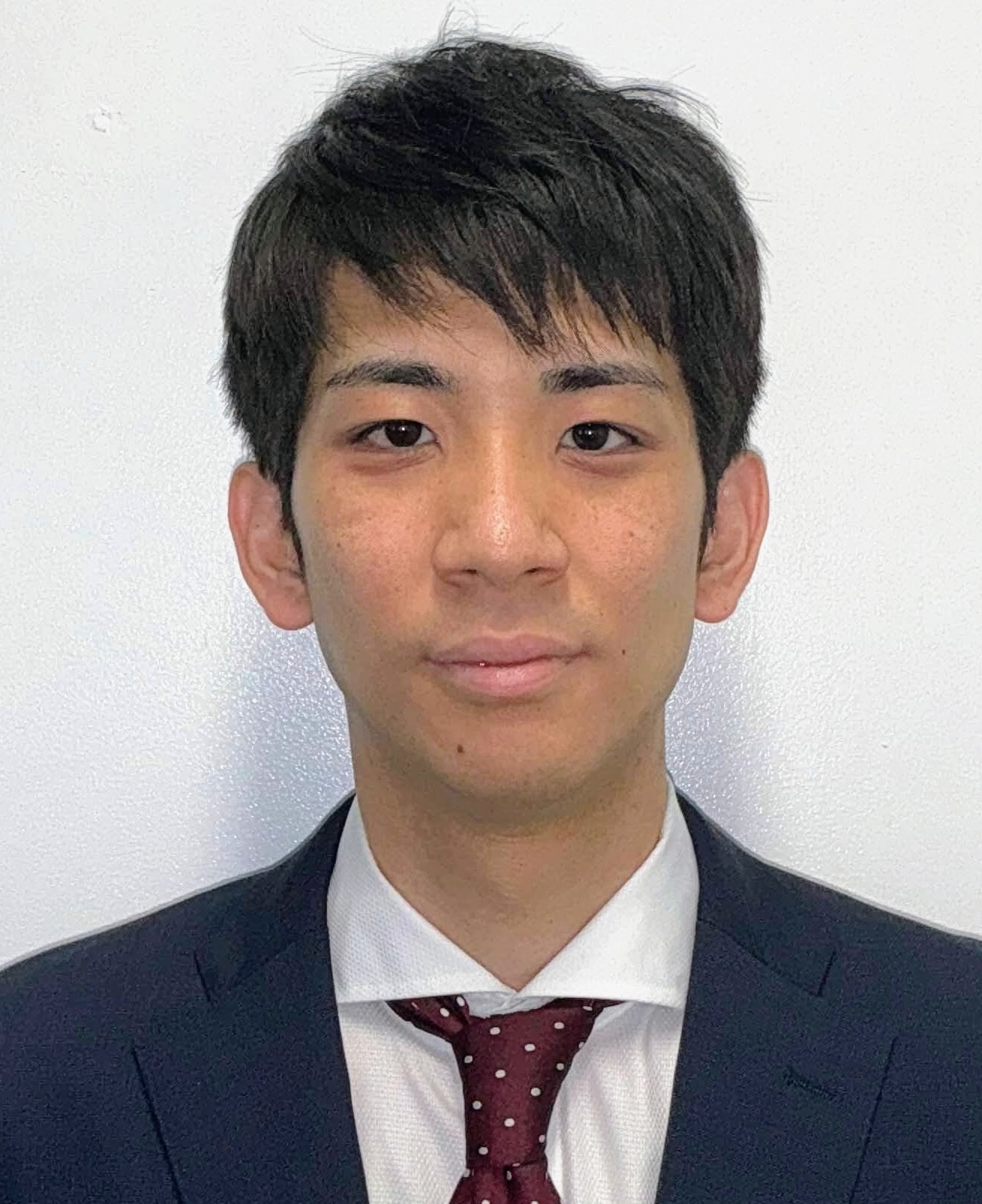}}]{Kenta Yokoe}
		(Member, IEEE) received the B.E. degree in mechanical engineering, the M.E. degree, and the Ph.D. degree in micro--nano mechanical science and engineering from Nagoya University, Nagoya, Japan, in 2021, 2023, and 2025, respectively.  He is currently an Assistant Professor at Nagoya University. His research interests include virtual reality, human interfaces, human--machine interaction, and haptic technology.
	\end{IEEEbiography}
	
	\begin{IEEEbiography}[{\includegraphics[width=1in,height=1.25in,clip,keepaspectratio]{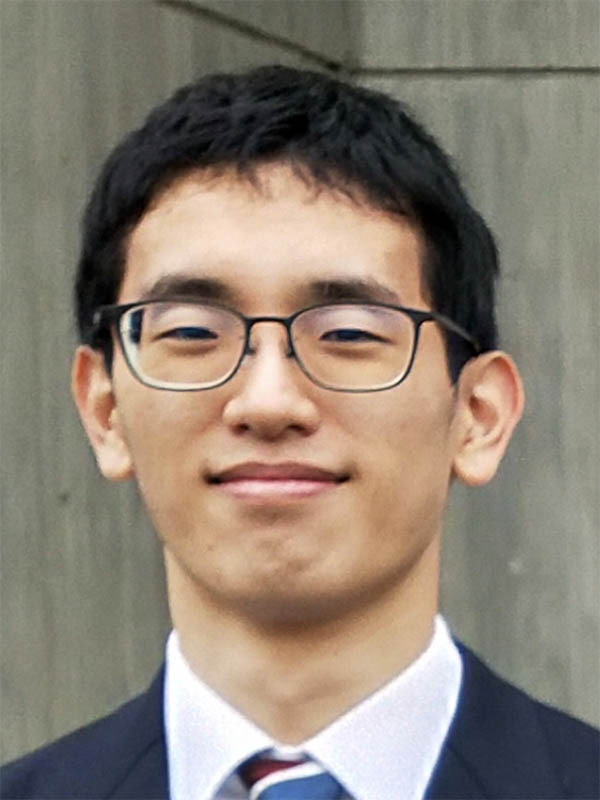}}]{Takuya Hara}
		received the B.E. degree in mechanical engineering, and M.E. degree in micro--nano mechanical science and engineering from Nagoya University, Nagoya, Japan, in 2021 and 2023, respectively. He is currently with DENSO Corporation.
	\end{IEEEbiography}
	
	\begin{IEEEbiography}[{\includegraphics[width=1in,height=1.25in,clip,keepaspectratio]{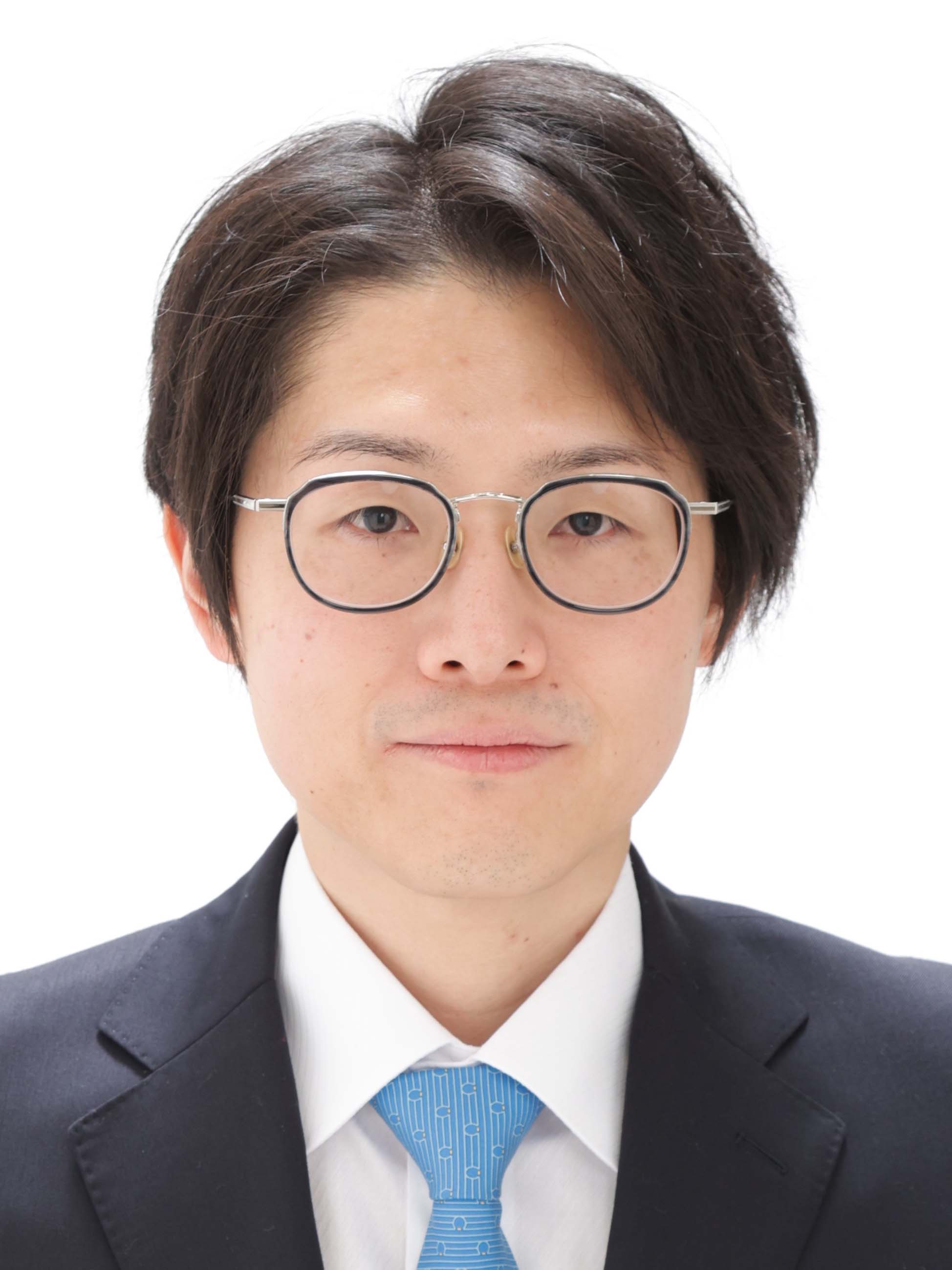}}]{Tadayoshi Aoyama}
		(Member, IEEE) received the B.E. degree in mechanical engineering, the M.E. degree in mechanical science and engineering, and the Ph.D. degree in micro--nano systems engineering from Nagoya University, Nagoya, Japan, in 2007, 2009, and 2012, respectively. He was an Assistant Professor with Hiroshima University, Higashihiroshima, Japan, during 2012--2017, and Nagoya University during 2017--2019. He then became an Associate Professor with Nagoya University during 2019--2024. From 2018 to 2022, he was a PRESTO Researcher with JST. He is currently a Professor with the Department of Mechanical Systems Engineering, Nagoya University. His research interests include macro--micro interaction, VR/AR and human interfaces, AI-based assistive technology, micromanipulation, and medical robotics. 
	\end{IEEEbiography}
	
	\EOD
\end{document}